\documentclass[prx,aps,twocolumn,superscriptaddress,floatfix,longbibliography]{revtex4-2}

\usepackage[separate-uncertainty = true,multi-part-units=single]{siunitx}
\usepackage{upgreek}
\usepackage{bm,amssymb,amsmath,amsthm,mathtools,xcolor,array}

\newcommand{\new}[1] {\textcolor{blue}{#1}}
\renewcommand{\new}[1] {#1}

 \newcommand{\alphar} {\alpha_\text{rev}}
 \newcommand{\alphat} {\alpha_\text{tum}}
 \newcommand{\alphaf} {\alpha_\text{fru}}
 
\begin{document}

\title{Preventing clustering of active particles in microchannels}

\author{Juan Pablo Carrillo-Mora}
\affiliation{Departamento de Física, FCFM, Universidad de Chile, Santiago, Chile}
\affiliation{Departament de Física de la Matèria Condensada, Universitat de Barcelona, Barcelona, Spain}
\affiliation{UBICS University of Barcelona Institute of Complex Systems, Barcelona, Spain}

\author{Moniellen Pires Monteiro}
\affiliation{Departamento de Física, FCFM, Universidad de Chile, Santiago, Chile}

\author{V.I. Marconi}
\affiliation{Facultad de Matemática, Astronomía, Física y Computación, Universidad Nacional de Córdoba and IFEG-CONICET, Córdoba, Argentina}

\author{Mar{\'\i}a Luisa Cordero}
\affiliation{Departamento de Física, FCFM, Universidad de Chile, Santiago, Chile}

\author{Ricardo Brito}
\affiliation{Departamento de Estructura de la Materia, Física Térmica y Electr{\'o}nica and GISC, Universidad Complutense de Madrid, Spain}

\author{Rodrigo Soto}
\affiliation{Departamento de Física, FCFM, Universidad de Chile, Santiago, Chile}

\begin{abstract}
The trajectories of microswimmers moving in narrow channels of widths comparable to their sizes are significantly altered when they encounter another microswimmer moving in the opposite direction. The consequence of these encounters is a delay in the progress of both swimmers, which can be conceptualized as an instantaneous effective backward displacement. Similarly, the modeling of tumble events in bacteria, which occur over a finite time, can be represented as an instantaneous effective displacement in addition to a change in direction. Such effective displacements can be incorporated directly into a kinetic theory for the partial densities of swimmers moving in the channel. The \new{linear} analysis of the resulting equation yields the critical density at which clusters emerge. The methodology is then 
applied to the case of soil bacteria moving in long channels of cross-section  1.8~${\upmu}$m $\times$
1.8~${\upmu}$m.  The tracking of the swimmers permits the straightforward acquisition of the effective displacements, which in turn allows the critical density (${\rho}_{\text{crit}}\simeq$ 0.10 bact/${\upmu}$m) to be predicted prior to cluster formation. The advantage of this proposed approach is that it does not necessitate the determination of an effective density-dependent speed, which is a requisite of the standard motility-induced phase separation theory.
\end{abstract}

\maketitle


Soft lithography microfabrication techniques are enlarging our knowledge of microscopic life~\cite{Weibel2007, Hol2014}. Through the fabrication of microdevices, structures and actuators with dimensions  comparable to the size of microorganisms, the microenvironment can be controlled with unprecedented detail, allowing the visualization and quantification of biological processes at the cellular, microorganismal and communal scales. These advances have led to deeper understanding of biological processes, such as motility and chemotaxis~\cite{Mao2003, Tokarova2021, carrillo2025damage}, signaling and quorum sensing~\cite{Park2003,Taylor2010, vanVliet2014}, cell division and propagation of phenotypes~\cite{Balaban2004,Mannik2009,Wang2010}, bacterial motility in porous and disordered media~\cite{Bhattacharjee2019, Scheidweiler2020, monteiro2025soil}, organisms interactions~\cite{Massalha2017, Arellano-Caicedo2021}, among many other examples. 
At the same time, understanding the physical behavior of microswimmers under confinement has enabled the design of devices for applications, such as bacterial sorting~\cite{DiLuzio2005, Galajda2007,wan2008rectification, berdakin2013a, berdakin2013b}, or to extract useful power from swimming bacteria~\cite{Hiratsuka2006}, based on narrow channels or funnels with apertures of only a few bacterial body sizes.

The drawback of confinement is that obstructions between individuals become relevant as devices reduce their size.   
Indeed, clusters may  emerge in quasi 1D microchannels when microswimmers can overpass, but still partially block for a while, because at an encounter they reduce their speed, leaving time for more particles to come and form the cluster. 
Clusters must be avoided, as once they appear are difficult to dissolve and, in the case of bacteria, they can quickly develop into biofilms. Notably,  effective quasi 1D dynamics also emerge in wide microfluidic devices as bacteria naturally tend to swim on the edges~\cite{DiLuzio2005, figueroa2015living} and moving 1D clusters appear~\cite{perez2024accumulation}. 
Active wetting, in which persistent agents remain long times in contact with walls, also gives rise to  quasi 1D dynamics~\cite{sepulveda2017wetting,sepulveda2018universality}.

The theoretical description for the clustering process in microchannels was first put forward by Tailleur and Cates~\cite{tailleur2008statistical}, assuming that there is an effective reduction in the swimming speed, $V_\text{eff}$, depending on the bacterial density $\rho$. They show that if $V_\text{eff}(\rho)$ decreases sufficiently rapidly with $\rho$, an instability develops. This mechanism of motility-induced phase separation (MIPS) has been extended to two and three dimensions~\cite{cates2015motility}.

Despite its relevance to predict the conditions that can trigger clustering, until recently there was no fully microscopic derivation of $V_\text{eff}$ in terms of the kinematic properties of the swimmers. In Ref.~\cite{soto2024kinetic}, we conceptualized the persistent encounters of active Brownian particles (ABP), in which they slide on each other, as collision events where particles are instantaneously displaced with respect to their free motion in the absence of a collision. With this kinetic approach, we were able to predict the density where MIPS takes place, with no need of introducing an effective velocity and only using microscopic parameters.

Extending this idea to the case of bacteria moving in microchannels, here we provide  a simple experimental method to be implemented at low density conditions, which gives the maximum swimmer density and channel length to avoid cluster formation. As in ABPs, there is no need to measure the effective velocity reduction as a function of density, as the standard MIPS theory requires.
Instead, the proposed method consists in identifying and quantifying only the bacterial motility and simple \textit{bacterial dynamical events} involving one (tumbles, temporary arrests, ...) or two bacteria (collisions). Importantly, the motility and these events must be measured in the channel under study and not in the bulk.

\section{Bacterial motility in microchannels}
\begin{figure}[htb] 
\begin{center}
\includegraphics[width=.8\columnwidth]{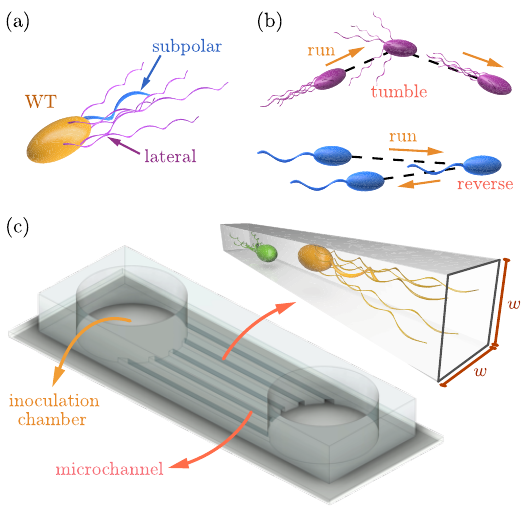}
\end{center}
\caption{
(a) Schematic representation of \textit{B.~diazoefficiens} WT bacterium with its two flagellar systems: lateral and subpolar. In a typical WT bacterial suspension, the average cell body size along its long axis is approximately \SI{1.5}{\micro\meter}, with an average aspect ratio of approximately 2:1. 
(b) Run-and-tumble and run-and-reverse swimming strategies.
(c) Scheme of the microfluidic device, which consists of two cylindrical reservoirs (inoculation chambers) connected by a set of microchannels arranged in parallel. The cross section of these channels is square with a side length of $w \approx \SI{1.8}{\micro\meter}$.
}
\label{fig.setup}
\end{figure}

To study the dynamics in microchannels, we use \textit{Bradyrhizobium diazoefficiens}, although the theory we present is of general applicability. These bacteria 
 move through soil pores to penetrate the roots of soybean plants, forming symbiotic nodules that
fix nitrogen  \cite{franche2009nitrogen,16quelas2016, 20garrido2019}. This symbiotic relationship enhances crop growth, thus \textit{B. diazoefficiens} 
serve as a biofertilizer and a sustainable complement to traditional inorganic fertilizers, 
which have potentially harmful  effects \cite{Pahalvi2021}.
The wild-type (WT) of
\textit{B.~diazoefficiens} has two flagellar systems \cite{19quelas2016}: the subpolar,
consisting of a long, single flagellum located near one pole of its body, and the lateral system, consisting of several longer and thinner flagella  along the  body (see Fig.~\ref{fig.setup}a). Studies about its motility  \cite{19quelas2016, monteiro2025soil} revealed that they  perform run-and-tumble swimming strategies, like \textit{Escherichia coli}  \cite{22grognot2021, 23sartori2018}, and also run-and-reverse strategies, as \textit{Pseudomonas aeruginosa}  \cite{23sartori2018, tian2022}.
In a \textit{tumble} event the bacterium changes its swimming direction by reorienting its body. In a \textit{reverse} event it swims backwards, without reorienting its body  (see Fig.~\ref{fig.setup}b).

 To study all the dynamical events in microchannels, we designed and fabricated  an array of parallel microchannels  (see Fig.~\ref{fig.setup}c) of square cross-section of width $w = \SI{1.8}{\micro\meter}$ \new{and \SI{2.25}{\milli\meter} in length}. The channels  were connected to two inoculation chambers, acting as reservoirs for a bacterial suspension.

Bacterial trajectories were recorded for subsequent analysis.  
The mean swimming speed  $v_m$ was measured  in intervals without tumbles, building a histogram (PDF) shown in  Fig.~\ref{fig.motility}. The mean and standard deviation (in absolute value) were $\langle v_m \rangle = \SI{6.1 \pm 2.9}{\micro\meter/\second}$ (obtained from 1270 intervals). For simplicity, velocity fluctuations will be neglected, and it will be assumed that bacteria either move to the left or right at the average speed, $\langle v_m\rangle$, denoted by $\pm V$.

\section{Bacterial dynamical events and kinetic description}

The description of the bacterial dynamics requires the use of the tools of kinetic theory 
to reflect the changes in position or velocities due to all possible  events. 
We  denote by $f_+(x,t)$ the density of bacteria at position $x$ at time $t$ moving with positive velocity, $+V$, and 
$f_-(x,t)$ when moving with negative velocity, $-V$, normalized such that the  bacterial density is $\rho(x,t)=f_+(x,t)+f_-(x,t)$.
Kinetic theory writes equations for the variation of such functions for each of the bacterial events that will be described below~\cite{liboff2003kinetic,soto2016kinetic}.

\subsection{Runs}

In the run phase,  bacteria swim in a straight line inside the microchannel at a constant speed. Then, a bacterium swimming with velocity $+V$ moves from $x$ to $x+V\Delta t$  in a  time $\Delta t$. In kinetic language: $f_+(x,t)=f_+(x+V\Delta t,t+\Delta t)$. This balance equation will be supplemented with a term  $J_+$, coming from tumbles, reverses and collision events. Making a Taylor expansion for $\Delta t\to 0$, the equation reads

\begin{align}
\frac{\partial f_{+}}{\partial t} +  V \frac{\partial f_{+ }}{\partial x}=J_+.
\label{eq.free2}
\end{align}
There is a similar equation for bacteria swimming in the left direction,  $f_-$, with the term $+V$  replaced by $-V$ .

\subsection{Rates of change of direction}
\begin{figure}[htb] 
\begin{center}
\includegraphics[width=.8\columnwidth]{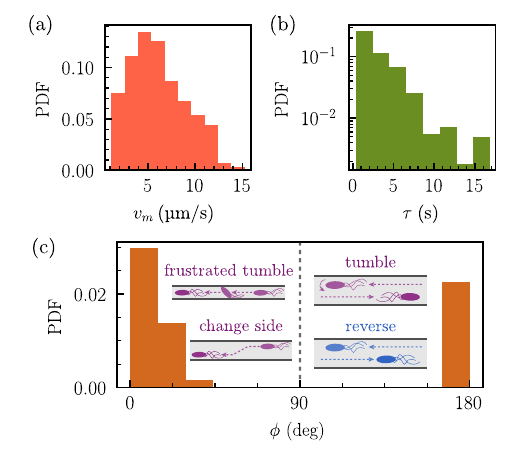}
\end{center}
    \caption{Probability distributions for (a) mean swimming speed $v_m$, (b) characteristic time between events of change of direction $\tau$ in semilog scale, and (c) angles of change of direction $\phi$, with the associated events.}
\label{fig.motility}
\end{figure}

From the analysis of the bacterial trajectories, we measure the  time between events of change of swimming direction, $\tau$, and the angle of change of direction in each of these events, $\phi$. 
Figure~\ref{fig.motility}b presents the PDFs of $\tau$, with average and standard deviation $\langle \tau \rangle = \SI{3.3 \pm 2.9}{\second}$ (obtained from 255 events).
The histogram presents an exponential distribution that, together with the standard deviation being very close to the mean, imply that they obey a Poisson distribution, with the total rate of change of direction being $\alpha_0= \SI{0.31\pm0.02}{\second^{-1}}$. 
This total rate  can be decomposed into partial rates by analyzing the angle of change of direction (see Fig.~\ref{fig.motility}c and Materials and Methods). The result  is $\alpha_\text{tum} = \SI{0.0523\pm0.0080}{\second^{-1}}$ for tumbles, $\alpha_\text{rev} = \SI{0.0061\pm0.0027}{\second^{-1}}$ for reversals, and $\alpha_\text{fru} = \SI{0.0462\pm0.0075}{\second^{-1}}$ for frustrated tumble events, described in detail below. The rest are  events in which a bacterium swimming close to one of the walls becomes misaligned and crosses the microchannel to swim aligned with the parallel wall~\cite{monteiro2025soil}. Such events, visible in the Supplementary Video 1, 
do not produce any noticeable effect in the 1D projected bacterial motion, and will not be considered in the present study.

\begin{figure*}[htb] 
\includegraphics[width=2.0\columnwidth]{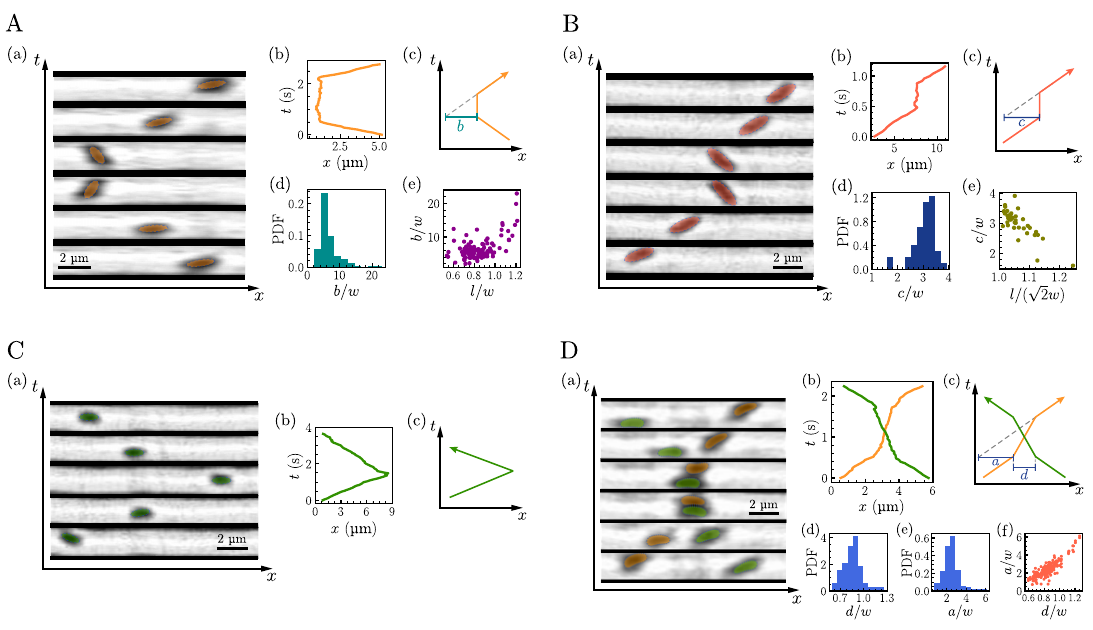}
    \caption{Possible events of a bacterium in a narrow channel. 
    (A) Tumble,
    (B) Frustrated Tumble,
    (C) Reverse,
    (D) Collision.
    Panels labeled  (a) are  image sequence of the  event (the time interval between images is $\Delta t \approx \SI{0.4}{\second}$). Full movies are shown in the Supplementary Information. 
    Panels (b) are experimental space-time diagrams of each event, obtained by particle tracking.
    Panels (c) are space-time representations of panels (b) with effective displacements defined [parameters $b$, $c$ and $a$ for tumbles (A), frustrated tumbles (B), and collisions (D)]. 
    Panels (Ad), (Bd), (Dd), and (De) are the probability distributions for the effective displacements, $b$, $c$, $d$, and $a$ respectively,  normalized by the width of the microchannel $w$. 
    Panel (Ae): Effective displacement $b$ versus the length of the bacterium $l$, both normalized with respect to $w$.
    Panel (Be): Normalized effective displacement $c$ normalized with respect to $w$ as a function of the ratio between the length of the bacterium $l$ and the maximum length of the bacterium that allows changes of direction with angles greater than \ang{90}, $\sqrt{2}w$.
    Panel (Df):  Effective displacement $a$ versus the extent of interactions $d$, both normalized with respect to $w$.}
\label{fig.tumble}
\end{figure*}

\subsection{Tumbles}
A tumble changes the swimming direction by reorienting the bacterium's body, a process that takes a finite time, as illustrated in  Fig.~\ref{fig.tumble}.Aa and Supplementary Video 2. 
Initially the bacterium  is moving to the left; then it performs the tumble event, during which it stands still for some time while reorienting its body,  and eventually moves to the right. The corresponding space-time diagram is shown in  Fig.~\ref{fig.tumble}.Ab.
The process is conceptually represented in Fig.~\ref{fig.tumble}.Ac, where the stopping time has associated a horizontal effective displacement $b$ to the left with respect to the trajectory the swimmer would have followed if the tumbling were instantaneous. 
The whole tumbling process is then modeled as an \textit{instantaneous} process in which the swimmer is displaced by $b$ \textit{and} changes direction. 
The PDF for the values of $b$ measured in the experiments is shown in Fig.~\ref{fig.tumble}.Ad, and its mean and standard deviation  are $\langle b \rangle = \SI{11.8 \pm 6.0}{\micro\meter}$ (obtained from 107 events), i.e. $\langle b/w \rangle = 6.5 \pm 3.3$ (see Materials and Methods for details).
Note that the difficulty for a bacterium to complete the tumbling process increases as its length   $l$  approaches the channel width $w$. 
The above translates into a rapid increase in the value of $b$ when  $l$  exceeds $w$ (see Fig.~\ref{fig.tumble}.Ae), while remaining approximately constant when $l < w$. 
The maximum length of a bacterium that allows tumbling with angles greater than $\ang{90}$ is $l\approx \sqrt{2}w$; larger bacteria cannot tumble. Such events are named {\em frustrated tumbles} and analyzed in a subsection below.

The tumbling event changes a bacterium with speed $-V$ into  a bacterium of speed $+V$ or viceversa, and then the contribution of the tumble event to Eq.~(\ref{eq.free2}) will include both $f_+$ and $f_-$. 
In the kinetic description, for a bacterium to appear at $x$ moving to the right after a tumble (as in Fig.~\ref{fig.tumble}.Ac), the bacterium moving to the left must have been located at $x+b$, such that after the effective displacement $b$ to the left it ends at $x$. With this,
the contribution to the kinetic equations of this event is:  

\begin{align}
J^{(\text{tum})}_{+}  =& \alpha_\text{tum}[f_-(x+b) - f_+(x)],  \\
J^{(\text{tum})}_{-} =& \alpha_\text{tum}[f_+(x-b) - f_-(x)].
\label{eq.tumble }
\end{align}
Note that in this formulation, the finite duration of the event is translated into a finite spatial displacement. Keeping the original delay in time would lead to a  non-local equation in time.

\subsection{Frustrated tumbles} 
Bacteria present size variability.
When the size of a bacterium, $l$, exceeds  $\sqrt 2 w$,  there is not enough space to execute a tumble, leading to a frustrated event. Such event is illustrated in Fig.~\ref{fig.tumble}.Ba (Supplementary Video 3),
where a bacterium moving to the right tries to turn, but ultimately continues in the same direction.
 The space-time diagram is represented in Fig.~\ref{fig.tumble}.Bb and the schematic one in Fig.~\ref{fig.tumble}.Bc where again, an effective displacement $c$ is defined. Similarly as for tumbles, this process is modeled as an instantaneous one, where the swimmer is displaced a distance $c$ opposite to its motion. 
The PDF for the measured values of $c$ normalized by  $w$ is shown in Fig.~\ref{fig.tumble}.Bd. The mean and standard deviation of the effective displacements (obtained from 42 events) are  $\langle c \rangle = \SI{5.46 \pm 0.78}{\micro\meter}$ (i.e. $\langle c/w \rangle = 3.03 \pm 0.43$).
In contrast with tumbles and collision events (described below),  in which  the length of effective displacement increases with $l$, in frustrated tumbles the effective displacements decreases with $l$ (see Fig.~\ref{fig.tumble}.Be).

For the kinetic description, no crossed terms appear in the fluxes $J_\pm$, as the particle does not change direction, but the net effect of the process is to displace the swimmer. This results in 

\begin{align}
J^{(\text{fru})}_{+}  =& \alpha_\text{fru}[f_+(x+c) - f_+(x)],  \\
J^{(\text{fru})}_{-} =& \alpha_\text{fru}[f_-(x-c)-f_-(x)].
\label{eq.reverse }
\end{align}

\subsection{Reverse tumbles} 
Reverse events, contrary to the previous cases, are instantaneous events where the bacteria change speed without changing their body orientation (see Fig.~\ref{fig.tumble}.C, and Supplementary Video 4).
A total of 12 of these events were identified. The kinetic description is identical to  tumbles, but there is no effective displacement, that is, $b=0$ for reverse events. 
Its contribution to the kinetic equations (\ref{eq.free2}) is: 

\begin{align}
J^{(\text{rev})}_{+}  =& \alpha_\text{rev}[f_-(x) - f_+(x)],  \\
J^{(\text{rev})}_{-} =& \alpha_\text{rev}[f_+(x) - f_-(x)].
\label{eq.tumble}
\end{align}

\subsection{Collisions}

Collisions are such that two bacteria, moving in opposite directions, meet inside the microchannel. 
Figure~\ref{fig.tumble}.Da shows the sequence (Supplementary Video 5) of such an event and Fig.~\ref{fig.tumble}.Db the space-time diagram. 
\new{Here, as an effect of lubrication, steric or other forces, bacteria reduce their speed while bypassing each other.}  
 Similarly to ABPS~\cite{soto2024kinetic}, the collision  produces a net displacement $a$ on each swimmer, schematized in Fig.~\ref{fig.tumble}.Dc.  In these quasi-1D trajectories, swimmers first meet at a distance $d$, which is related to the size of the bacteria, $l$. At the end of the interaction, they are effectively displaced a distance $a$. 
\new{The trackings do not show evidence of bacteria interacting at long distances, which is consistent with the effectively reduction in range of the hydrodynamic interaction under strong confinement~\cite{Liron_Shahar_1978}.} 
 
The distributions for $d$ and $a$ are shown in Figs.~\ref{fig.tumble}.Dd-e, normalized by the microchannel width $w$. The average and standard deviation (106 collision events) for the extent of interactions are $\langle d \rangle = \SI{1.54 \pm 0.22}{\micro\meter}$ (i.e. $\langle d/w \rangle = 0.86 \pm 0.12$) and for the effective displacements are $\langle a \rangle = \SI{4.5 \pm 1.6}{\micro\meter}$ 
(i.e. $\langle a/w \rangle = 2.47 \pm 0.86$). 
 These variables are correlated, Fig.~\ref{fig.tumble}.Df, with a positive covariance $\sigma_{ad} = \big\langle (a-\langle a\rangle)(d-\langle d\rangle) \big\rangle=\SI{0.29}{\micro\meter^2}=0.027w^2$, meaning that the effective displacements increase with the extent of interactions. Since $d$ is proportional to the size of bacteria, the diversity of measured values for the extent of interactions is due to the variability in the bacterial size. 
 Higher values of $d/w$ increase the interaction time, therefore increasing $a$.

The collision term in the kinetic equation requires that {\em two} particles meet with opposite velocities. The probability of the occurrence is given by the product of $f_+$ and $f_-$ and the relative speed $2V$, resulting in: 
\begin{align}
\!J^{(\text{col})}_{+}\! &=\! 2V\left[f_+(x+a)f_-(x+a+d) - f_+(x)f_-(x+d) \right], \label{eq.collision.p}\\
\!J^{(\text{col})}_{-}\! &=\! 2V\left[f_-(x-a)f_+(x-a-d) - f_-(x)f_+(x-d) \right]. \label{eq.collision.m}
\end{align}
In kinetic theory, such terms are called {\em gain and loss} terms~\cite{liboff2003kinetic,soto2016kinetic}.  
In Eq.~\eqref{eq.collision.p}, a swimmer moving to the right is lost at $x$ after colliding with one moving to the left located at $x+d$ and to gain a swimmer moving to the right at $x$, the colliding swimmers should be at $x+a$ and $x+a+d$ (see Fig.~\ref{fig.tumble}.Dc). 
\new{Note that this description assumes that no tumbling or reverse is produced during a collision, which is correct in view of the small values of $\alpha$; the probability of such concurrent event is $\alpha\langle a\rangle/V$.} 

\section{Kinetic equation}
Finally, the kinetic equations for the motion of the bacteria are obtained by the addition of all the terms above:

\begin{align}
\frac{ \partial f_+}{\partial t} + V \frac{\partial f_+}{\partial x} =J^{(\text{tum})}_{+} + J^{(\text{fru})}_{+} + J^{(\text{rev})}_{+} + J^{(\text{col})}_{+},
\label{eq.Boltzmann}
 \end{align}
and  similarly for $f_-$.

\subsection{Linear stability analysis and critical density}
Linear analysis of Eq.~(\ref{eq.Boltzmann}) starts by  probing solutions around the homogeneous state of the form: 

\begin{align}
f_{\pm}(x,t) &= \rho/2 + g_{\pm} e^{\lambda t + ikx},
\label{eq.modes}
\end{align}
where $k$ is the wavenumber, the eigenvalue  $\lambda$ is the growth rate of the mode, and the quantities $g_{\pm}$ are the amplitudes of the modes. 
The stability of the homogeneous solution is determined by the sign of the real  part of the eigenvalues.  Substitution of the ansatz~(\ref{eq.modes}) in the kinetic Eq.~(\ref{eq.Boltzmann}) yields: 

\begin{align}
\mathbb{A} \begin{pmatrix}
g_+\\ g_-
\end{pmatrix}  = \lambda \begin{pmatrix}
g_+\\ g_-
\end{pmatrix},
\label{eq:eigenvalueproblem}
\end{align}
 where the lineal collision operator, $\mathbb{A}$, is obtained as:  
 
\begin{align}
\mathbb{A}=&
ikV\begin{bmatrix}
-1&0\\
\phantom{-}0 & 1\\
\end{bmatrix} +
\alpha_\text{tum}\begin{bmatrix}
-1 &  e^{i kb}\\
e^{-i kb} & -1\\
\end{bmatrix}\nonumber \\
&+
\alpha_\text{fru}\begin{bmatrix}
e^{ikc}-1 & 0\\
0 &e^{-ikc}-1\\
\end{bmatrix} 
+\alpha_\text{rev}\begin{bmatrix}
-1&\phantom{-}1\\
\phantom{-}1 & -1\\
\end{bmatrix}
 \nonumber\\
&+ \rho V\begin{bmatrix}
e^{ika}-1&
e^{ik(d+a)}-e^{ikd}\\
e^{-ik(d+a)}-e^{-ikd} & 
e^{-ika}-1\\
\end{bmatrix}.
\label{eq.Amatrix}
\end{align}
Some properties of $\mathbb{A}$ and its averages are deferred to the Materials and Methods.  Note that reverse events are unimportant  in {\em B.\,diazoefficiens} and absent in {\em E.\,coli}, so they are neglected.  The $k$ expansion of the relevant eigenvalue is:

\begin{multline}
\lambda=-\bigg\{
\alpha_\text{tum}  
\alpha_\text{fru}\sigma_b^2\langle c^2\rangle
-2\rho  V \left(\langle a^2\rangle +\langle ad\rangle -\langle a\rangle \langle b \rangle \right)\\
\,\,+\frac{\left(V-\alpha_\text{fru}\langle c \rangle\right)\left(V-\alpha_\text{fru}\langle c \rangle-2 V \langle a\rangle \rho\right)}{\alpha_\text{tum}} 
\bigg\} \frac{k^2}{2} + {\cal O}(k^4).
\label{eq.lambda2}
\end{multline}

The stability criterium for the density is determined by setting the coefficient of the term $k^2$ to zero:
\begin{align}
\rho_{\text{crit}}= \frac{1}{2}\frac{(1-\alpha_\text{fru}\langle c\rangle /V)^2+\alpha_\text{tum}^2\sigma_b ^2/V^2+\alpha_\text{tum}\alpha_\text{fru}\langle c^2\rangle}{\langle a\rangle-\alpha_\text{tum}\left(\langle a^2\rangle +\langle ad\rangle -\langle a\rangle \langle b \rangle \right)/V -\alpha_\text{fru}\langle a\rangle\langle c\rangle/V}. \label{eq.rhocrit}
 \end{align}
For bacterial density above $\rho_{\text{crit}}$, the system is linearly unstable and bacteria will cluster and clog the microchannel. On the contrary, for $\rho<\rho_{\text{crit}}$ no clusters appear. 
Evaluation of Eq.~(\ref{eq.rhocrit}) with the values measured for the bacteria (see Materials and Methods) gives

\begin{align}
\rho_{\text{crit}}\simeq \SI{0.10}{bact/\micro\meter}.
\end{align}
Recalling that the typical size of a bacterium is $\langle d\rangle \simeq \SI{1.54}{\micro\meter}$, this means that bacteria occupy around 15\% of the  available space. This linear density translates into a volumetric density of \SI{0.031}{bact/\micro\meter^3} inside the channel. Noe that this density may not correspond to the density in the inoculation chambers~\cite{DiLuzio2005,figueroa2015living,figueroa2020coli,Galajda2007,wan2008rectification}.

Moreover, the effect of tumbling and frustrated tumbling events on the critical density is small, with dimensionless factors  $\alpha_\text{tum} b/V\simeq 0.10$ and $\alpha_\text{fru} c/V\simeq 0.04$. Keeping just the contributions of collisions it reads,      

\begin{align}
\rho_{\text{crit}} \simeq \frac{1}{2\langle a\rangle}\simeq\SI{0.11}{bact/\micro\meter}, \label{eq.rhocrit.simple}
\end{align}
very close to the full result, confirming that collisions are the main dynamical ingredient as theorized in the MIPS formulation~\cite{tailleur2008statistical,cates2015motility}. 
\new{Speed variability expresses as $V$ being a stochastic variable in $\mathbb{A}$, implying that $V$ now enters into the averages and  correlations appear in Eq.~\eqref{eq.lambda2}. Nevertheless, the dominant contribution for the critical density [Eq.~\eqref{eq.rhocrit.simple}] is insensitive to speed variability, validating the choice for neglecting these effects in the analysis. }

\begin{figure}[htb]  
\includegraphics[width=.9\columnwidth]{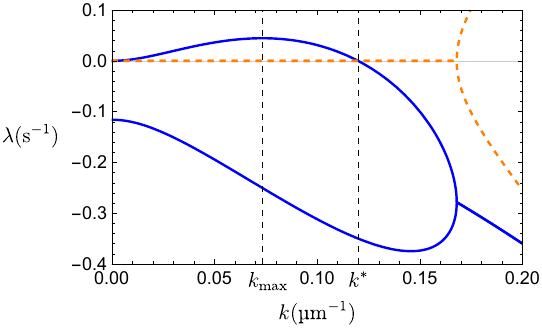} 
\caption{Example of the spectrum for an unstable case ($\rho=\SI{0.115}{bact/\micro\meter}>\rho_{\text{crit}}$). Solid (dashed) lines are the real (imaginary) parts of the eigenvalues. 
The vertical dashed lines give the most 
unstable wavenumber $k_\text{max}\simeq\SI{0.073}{\micro\meter^{-1}}$ and 
the  threshold wavenumber $k^*\simeq\SI{0.12}{\micro\meter^{-1}}$. } 
\label{fig.spectrum} 
\end{figure}

\subsection{Analysis}

 Equation~(\ref{eq.rhocrit}) demonstrates that collisions are essential to produce clustering: if  $\langle a\rangle=0$ (implying that $\sigma^2_a=\sigma_{ad}=0$ as $a\geq 0$) the critical density is infinite. In the same spirit, small swimmers ($d$ small) are harder to cluster. 
 
 Analysis of Eqs.~\eqref{eq.lambda2} and \eqref{eq.rhocrit} shows that processes that slow the motion of swimmers facilitate clustering in the spirit of Ref.\cite{tailleur2008statistical}. In other words, the longer the particles remain stopped (while tumbling) the easier to cluster. This can happen by many  ($\alpha_\text{tum}$ large) or slow ($b$ large) tumbles or many and slow frustrated tumbles ($\alpha_\text{fru}$ and/or $c$ large).
On the contrary, fast tumbles ($\langle b \rangle \to 0$) or very few turns (small $\alpha_\text{tum}$) require a higher density to produce clusters. 
Finally, the variability in the parameters $a$, $b$ and $c$, and positive correlation between $a$ and $d$ make the system more stable.

  As modes with positive eigenvalues grow with time, there is a fastest growing mode in Fig.~\ref{fig.spectrum}, associated with a wavenumber $k_\text{max}$. Then, \new{when the linear regime is satisfied (at short times)}  the mode with $k=k_\text{max}$ is selected, creating space oscillations of density with a characteristic wavelength  $\xi\simeq 2\pi / k_\text{max}$.  
  \new{At later times the nonlinear terms in the Eqs.(8) and (9) take over and produce a saturation of the functions $f_{\pm}$ at the crests and a vanishing density at the troughs. Then  
 the distance between clusters is approximately $\xi$. Moreover,} if the channel length $L$ is shorter  than the threshold $L^*=2\pi/k^*$, the system is overall stable as it cannot contain an unstable mode. The threshold $k^*$ is obtained by the condition $\lambda(k^*)=0$ or  $\det(\mathbb{A})=0$ and depends on $\rho$.
For example, with $\rho\simeq\SI{0.115}{bact/\micro\meter}$, slightly above the critical value, it results in $L^*\simeq\SI{52}{\micro\meter}$. Channels of length $L<L^*$ will not clog, whereas those with $L>L^*$  develop clusters and eventually clog. 

\new{Based on the kinetic equation, it is also possible to study the motion of a tagged swimmer. The result, for long times, is a Fokker--Planck equation, which describes the effective diffusive motion of the swimmer due to the interactions with the others~\cite{SotoPoF2025}.}

\section{Discussion}
Understanding the dynamics of microorganisms under confinement in microchannels is crucial for advancing future applications of microfluidic devices, which offer a wide range of potential uses.
Here we present a kinetic theory to describe all possible events that a bacterium can undergo: tumbles, reverses and collisions with other bacteria. Moreover, the narrow aspect of the microchannel can produce tumbles that are not completed, so-called frustrated tumbles. Each of these events is represented by a contribution to the kinetic equation, where it is fairly simple to carry out a linear stability analysis. 
That analysis yields a critical density and maximum length of microchannel  that will not clog, helping in the design of actual devices, avoiding clogging or biofilm formation. 

\new{The present study considered bacteria under strong confinement. Increasing the channel widths makes easier for the swimmers to bypass each other, translating into smaller effective displacements $a$, which, in turn, increases the critical density to observe clustering.  Preliminary experiments show that for $w=\SI{3}{\micro\meter}$,  $\langle a\rangle = \SI{1.15 \pm 0.94}{\micro\meter}$ ($N=26$), and for $w=\SI{4}{\micro\meter}$, the effective displacements are undetectable.
On the contrary, confining even further the swimmers will make more difficult for them to bypass each other, resulting in a clusters being formed more easily. 
It is also expected that the displacements $c$ for frustrated tumbles depend on the channel width. Nevertheless the analysis may be more complex, as for example, it has also been observed that higher confinements result in a reduction of the swimming speed~\cite{monteiro2025soil}, and specific measurements or modeling of swimmers under confinement with dynamical aspects of the bacterial motion as those considered in Ref.~\cite{berdakin2013a} are necessary. Also, experiments with \textit{E.~coli} indicate that the rate of successful tumbling events at walls is reduced~\cite{Stocker2014,Clement2022}, and similar mechanics can take place under confinement implying that the tumbling rate can depend on the channel width.}

\new{The measurement of the parameters were made in the bulk of the channels. Near the channel openings the parameters may change. Also, at high bacterial densities, quorum sensing responses may also change the tumbling rates. These effects, once have been experimentally or numerically characterized can be directly included into the kinetic equation.} Moreover, the theory can easily be generalized to other bacterial species more than \textit{B.~diazoefficiens}. It is hoped that these results will stimulate further experiments to confirm our findings, in particular our predictions for the critical channel length and cluster sizes.


\acknowledgments
{V.I.M. is thankful to Prof.~A.R.~Lodeiro for inspiration and for his knowledge and generosity on soil bacteria. This research was supported by the Fondecyt Grant No.~1220536 and Millennium Science Initiative Program NCN19\_170 of ANID. J.P.C.M. acknowledges funding from ANID Beca de Magíster Nacional No. 22221639. M.P.M. acknowledges the postdoctoral Fondecyt Grant No.~3190637. V.I.M. acknowledges support from Grants: SeCyT-UNC: 33620230100298CB; FONCyT: PICT-2020-SERIEA-02931; CONICET:PIP-2023-11220220100509CO. The work of R.B. is supported from Grant Numbers PID2020-113455GB-I00 and PID2023-147067NB-I00. Fabrication of microfluidic devices was possible thanks to ANID Fondequip grants Nos.\,EQM140055 and EQM180009.}


\appendix
\ \\ 
\section{\uppercase{Materials and methods}} \label{app.methods}
\subsection{Bacterial culture}
For routine use of the wild-type strain of \textit{B. diazoefficiens} USDA 110, bacterial stocks were maintained at 4°C in solid yeast extract mannitol-agar medium (YEM agarized 1.5\%) \cite{58dardis2021}, which were renewed every three months. For the experiments, from a single colony in the agar plates, cultures in liquid growth medium HMY-arabinose \cite{18mengucci2020} were initiated and grown, to late log phase (reaching an optical density at 600 nm wavelength \cite{39beal2020} OD$_{600}$ = 3.0 $\pm$ 0.1) at a temperature of 28°C and 180 rpm of agitation. The bacteria were then recultured by diluting an aliquot of the first culture in 10 mL of the same medium (HMY-arabinose), setting the initial OD$_{600}$ to 0.1. This reculture was grown at the same temperature and agitation until it reached OD$_{600}$ = 1.0 $\pm$ 0.1. Finally, a 1:100 dilution of the reculture in HM-arabinose minimal medium was kept at rest without agitation at 28°C for six hours before inoculation into the microfluidic device for an optimal motility.

\subsection{Microchannel fabrication}
The fabrication process of the microfluidic device was divided into two parts: the first part consisted of fabricating a mold using maskless optical lithography \cite{menon2005} on a silicon wafer with  SU-8 photoresist (Gersteltec Sarl). Then, using standard soft lithography techniques \cite{qin2010}, the second fabrication step consisted of generating replicas of the microchannel in PDMS (polydimethylsiloxane), which, once cured, were cut and removed from the mold using a scalpel. Followed, the holes for the inoculation chambers located at the ends of the microchannels were pierced with a 1.5 mm-diameter biopsy punch. Finally, the fabrication of the microfluidic device was completed by bonding the PDMS block to a glass slide by oxygen plasma activation \cite{50xiong2014}.

\subsection{Inoculation and data acquisition}
For a short time after oxygen plasma activation of the PDMS surface, the grouping of polar molecules on its surface makes it hydrophilic, which facilitates its wetting \cite{50xiong2014, hyunakim2011}. Thus, immediately after PDMS and glass bonding, the inoculation chambers at the ends of the microchannels were filled using a micropipette with a solution of minimal medium HM-arabinose and PVP-40 (Polyvinylpyrrolidone-40, Sigma Aldrich) at 0.05\% w/v. As the surface in the device was still hydrophilic, all microchannels were flooded by capillarity, without leaving air bubbles inside. By using PVP-40 in solution with minimal medium, 
adhesion of the bacteria to the walls and self-agglutination were prevented~\cite{51jiang2023}.

Once all the microchannels of the device were flooded and after the resting period of the bacterial suspension, the bacteria were inoculated into the reservoirs. Pressure differences between the ends of the channels, which can cause undesired flows, were avoided. For that, the remaining fluid in the chambers from the previous device filling procedure was removed, and both chambers were filled again with the same volume of bacterial suspension in minimum medium HM-arabinose and PVP-40 at 0.05\% w/v. Once both chambers were filled, they were sealed with Parafilm.

The swimming of the bacteria in the microchannels was observed with a Nikon Eclipse TS100 microscope in bright field, a 40$\times$ magnification objective and an Andor Zyla sCMOS camera. Videos were recorded at 50 fps (frames per second), with an extension of 1000 frames (20 seconds) and a resolution of 
1024 $\times$ 1024 px$^2$. 
Calibration for the 40$\times$ magnification objective indicates a ratio of \SI{6.24}{px/\micro\meter}, this implies that the imaged regions had an area of 
164 $\times$ 164 \si{\micro\meter^2}.

\subsection{Data analysis}
The videos obtained from the experiments were processed with Fiji (ImageJ) software \cite{43schindelin2012} to enhance their brightness and contrast, and then analyzed with our in-house developed software Biotracker  \cite{44sanchez2016}, specially developed to obtain the trajectories of the bacteria and to detect and to quantify automatically their change of direction events. 

\subsection{Classification of events of change of direction} 
As observed in Fig.~\ref{fig.motility}c, approximately 67\% of the changes in direction exhibited angles $\phi$ in the range $[\ang{0}, \ang{45}]$, which were primarily due to tumbling events in which a bacterium swimming close to one of the walls became misaligned and crossed the microchannel to swim aligned with the parallel wall. Also in this range of angles were the frustrated tumble events, which were identified because the swimmer stoped for a finite time. The remaining 33\% of the changes of direction had angles in the range $[\ang{165},\ang{180}]$ and corresponded to reverse and tumbling events with angles $\sim \ang{180}$. Of those, tumblings were the cases in which the body rotated and reversals those in which there was a change of swimming direction without rotation of the body. The ratio between these two types of events were 1:9, i.e., reversals constituted approximately 11\% of the change-of-direction events with $\phi \sim \ang{180}$. In total, of the 255 events, 43 were tumbles, 5 reversals, and 38 frustrated tumbles. 

\subsection{Bacterial length measurement} 
Experimentally the camera records a  2D projection of the microchannel, overlaying the images in the direction perpendicular to the viewing plane, as the optical system has a depth of field similar to the depth of the channel, which allowed keeping in focus all the bacteria inside the channel. Thus, the measurement of the bacterial length $l$ were performed by averaging the length of the major axis of an ellipse fitted to the binarized image of the bacteria in the channel over the full extent of the frames in which the bacteria were kept within the observation region. 

\subsection{Determination of the effective displacements}
Considering that a dynamic event (tumble, frustrated tumble, or collision) occurs in the time interval between $t_i$ and $t_f$, the trajectories in the intervals \textit{I before} ($t < t_i$) and \textit{II after} ($t > t_f$) the event can be described by two straight lines of the form $x_{I, II}(t) = m_{I, II} t + n_{I, II}$. Thus, fitting the model $x_{I, II}(t)$ to the experimentally measured trajectory sections yielded the coefficients $m_{I, II}$ and $n_{I, II}$, from which the effective displacements associated with the different dynamic events were determined as: 
$a, b, c = \left| (m_{I}-m_{II}) t_i + n_{I} - n_{II}  \right|$.
Finally, for collision events it is required to record the trajectories of the two colliding bacteria: 
$x_{I, II}^\text{LR}(t)$ associated with the bacterium moving from left to right and $x_{I, II}^\text{RL}(t)$ associated with the bacterium moving in the opposite direction. Then, the extent of interactions was obtained as $d = \left(m_I^\text{RL} - m_I^\text{LR}\right)t_i + n_I^\text{RL} - n_I^\text{LR}$.

\subsection{Computation of the determinant and spectrum of the linear collision operator $\mathbb{A}$} 

The eigenvalues of $\mathbb{A}$ matrix in Eq.~\eqref{eq.Boltzmann} determine the stability of the solution  presented in Eqs.~(\ref{eq.modes}). If the real parts of all eigenvalues are negative, the exponentials in Eqs.~(\ref{eq.modes}) tend to zero, recovering the  homogeneous solution, $f_\pm\to \rho/2$. On the contrary, if at least one of the eigenvalues has a positive real part, the time evolution departs from the homogeneous solution, leading to linear inhomogeneities in the system and eventually nonlinear clustering. 
Both the trace and determinant of $\mathbb{A}$ are real, implying that the eigenvalues $\lambda$ are either real or a pair of complex conjugates. 
Particle conservation implies that one eigenvalue (the density mode) vanishes for $k=0$. The other one is negative and takes the value $-2(\alpha_\text{tum}+\alpha_\text{rev})$ when $k\to 0$. 
Finally, the eigenvalues of $\mathbb{A}$ only depend on even powers of $k$.

To compute  the eigenvalues, averages over the fluctuating quantities were taken. 
Analytically, for small $k$ the averages of the exponential functions were computed by using the cumulant expansion 
$\langle e^{ika}\rangle \approx e^{i k \langle a \rangle - \frac{1}{2}k^2 \sigma_a^2 + {\cal O}(k^3)}$, where $\sigma_a^2$ is the variance of $a$, and analogously for the other variables. Moreover, in the collision matrix the joint cumulant expansion was needed: $\langle e^{ik(a+d)}\rangle \approx e^{i k ( \langle a \rangle + \langle d \rangle )- \frac{1}{2}k^2 (\sigma_a^2 +\sigma_d^2 + 2 \sigma_{ad})+{\cal O}(k^3)}$, where $\sigma_{ad}$ is the covariance given before. A summary of the relevant averages, variances and covariances used in the evaluation of $\mathbb{A}$ for small $k$ are given in the Supplementary information. 
 
To obtain the full spectrum, for each value of $k$, the matrix elements of $\mathbb{A}$ in Eq.~(\ref{eq.Amatrix}) were obtained explicitly by computing numerically the average of the different exponentials over the set of experimental measurements of $a$, $b$, $c$, and $d$. Together with the numerical values of the rates and average velocity, $\mathbb{A}$ was obtained, from which the evaluation of the  determinant and spectrum were direct.


%


\clearpage

\large{\uppercase{Extended Data table:}}

\large{\textbf{Summary of measured values}} \label{app.summary}

\begin{table}[htb]
\centering
\begin{tabular}{lll}
\hline
\hline
Event & Parameter & Value\\
\hline
Tumbles & $\alphat$ &   \SI{0.052\pm0.008}{\second^{-1}}\\
             & $\langle b\rangle$ & \SI{11.8}{\micro\meter}\\ 
             & $\langle b^2\rangle$ & \SI{174}{\micro\meter^2}\\
             & $\sigma_b$ & \SI{6.0}{\micro\meter}              \\
\hline
Frustrated Tumbles & $\alphaf$ & \SI{0.046\pm0.007}{\second^{-1}}\\
             & $\langle c\rangle$ & \SI{5.46}{\micro\meter}\\ 
             & $\langle c^2\rangle$ & \SI{30.4}{\micro\meter^2} \\
             & $\sigma_c$ & \SI{0.78}{\micro\meter}              \\

\hline
Reverse Tumbles & $\alphar$ & \SI{0.006\pm0.003}{\second^{-1}}\\

\hline
Collisions & $V$ & \SI{6.1 \pm 2.9}{\micro\meter/\second}\\
             & $\langle a\rangle$ & \SI{4.45}{\micro\meter}\\ 
             & $\langle a^2\rangle$ & \SI{22.2}{\micro\meter^2} \\
             & $\sigma_a$ & \SI{1.55}{\micro\meter}              \\
	     & $\langle d\rangle$ & \SI{1.54}{\micro\meter}\\ 
             & $\langle d^2\rangle$ & \SI{2.42}{\micro\meter^2}\\
             & $\sigma_d$ & \SI{0.22}{\micro\meter}              \\
 	     & $\langle a d\rangle$ & \SI{7.14}{\micro\meter^2} \\ 
	     & $\sigma_{ad}$ & \SI{0.29}{\micro\meter^2}            \\
\hline
\hline
\end{tabular}
\caption{Summary of the rates and the mean, variance and standard deviations of the relevant parameters for all bacterial events, measured in the channel.}
\end{table}

\large{\uppercase{Supplementary Videos:}}

\begin{description}
\item[Supplementary Video 1]

{Example of a bacterium changing side, presenting no noticeable stopping phase or change of speed. The video, recorded at 50 frames per second, is reproduced twice, first at full speed and then slowed to 0.5x.  The format of the counter on the top-left corner is  hh:mm:ss:ff, with hh hours, mm minutes, ss seconds, and ff frames, the later running from 0 to 49.}

\item[Supplementary Video 2]

{Example of a bacterium presenting a tumble event. Video format and notation are the same as for Supplementary Video 1.}

\item[Supplementary Video 3]

{Example of a bacterium presenting a frustrated tumble event. Video format and notation are the same as for Supplementary Video 1.}

\item[Supplementary Video 4]
{Example of a bacterium presenting a reverse event. Video format and notation are the same as for Supplementary Video 1.}

\item[Supplementary Video 5]
{Example of a bacterium presenting a collision event. Video format and  notation are the same as for Supplementary Video 1.}

\end{description}

\end{document}